\documentclass{optica-article}

\journal{opticajournal} 

\articletype{Research Article}

\usepackage{bm}

\begin{document}

\title{Recovery-Oriented Optical System Design Based on Fisher Information}

\author{Zhuang Qian,\authormark{1} Yan Mo,\authormark{2} Hao Tan,\authormark{1} Yu Chen,\authormark{1} Haibo Wang,\authormark{2} 
Weining Guo,\authormark{1} Yijie Ni,\authormark{1} and Donglin Ma\authormark{1,2,*}}

\address{\authormark{1}School of Optical and Electronic Information, Huazhong University of Science and Technology, Wuhan, Hubei 430074, China\\
\authormark{2}MOE Key Laboratory of Fundamental Physical Quantities Measurement \& Hubei Key Laboratory of
Gravitation and Quantum Physics, PGMF and School of Physics, Huazhong University of Science and
Technology, Wuhan 430074, China\\
}

\email{\authormark{*}madonglin@hust.edu.cn} 


\begin{abstract*} 
Traditional optical system design for imaging applications has long been guided by image quality metrics such as geometrical RMS spot size, wavefront error, and modulation transfer function (MTF). Although these metrics effectively characterize direct imaging performance, they do not necessarily indicate optimal behavior when image restoration is subsequently applied. In modern imaging pipelines, however, optical systems are increasingly combined with physics-based or learning-based restoration methods, making the information transfer properties of the optical system critically important. In this work, we propose a recovery-oriented optical design approach that explicitly links optical system optimization with image restoration performance through an estimation-theoretic framework based on Fisher information. A new design criterion is adopted to ensure that the information encoded by the optical system can be reliably recovered during image restoration. To validate the proposed approach, two dual separated aspheric lenses are designed under identical configurations using conventional RMS-based optimization and the proposed recovery-oriented optimization, respectively. Although the RMS-optimized lens exhibits superior direct imaging quality, it underperforms after image restoration compared to the recovery-oriented design. These results demonstrate that recovery-oriented optical design provides a physically interpretable and effective extension of traditional optical design criteria for imaging systems operating within modern computational imaging pipelines.

\end{abstract*}

\section{Introduction}
Traditionally, the design of imaging optical systems has been centered on geometric image quality metrics such as the Root Mean Square (RMS) spot size, wavefront error, and the modulation transfer function (MTF)\cite{Rolland2021Freeform,Malacara2013Handbook,Smith2005Modern,Sasian2013Introduction,Yang2021Freeform}. These metrics provide intuitive physical interpretations and have proven highly effective in guiding the development of optical systems intended for direct imaging. For applications where the final image is obtained directly at the sensor, minimizing geometric aberrations remains a robust and well-established design strategy\cite{Jeong2006Low,Wu2017Algorithm,Qian2025Aberration,Qian2023Design,Wei2019Design,Cheng2009Design,Yang2017Automated,Bauer2024All,Zhuang2022Freeform}.

In many modern imaging systems, however, the optical image captured by the sensor is merely an intermediate state rather than the final output\cite{Mait2018Computational,Wetzstein2012Computational}. Instead, it is often subjected to digital restoration--such as deconvolution, inverse filtering, or learning-based enhancement--to mitigate optical blur and noise. As a result, optical image formation and subsequent computational processing are increasingly treated as a coupled system, where the specific aberrations of the optical design dictate the limits of computational recovery. In this context, the objective of optical design shifts from achieving immediate visual sharpness to optimizing the information capacity and its distribution across spatial modes for subsequent reconstruction.

This observation underscores a fundamental limitation of traditional RMS-based optical design when integrated with image restoration pipelines. While RMS spot size effectively reflects the spatial concentration of rays from a single object point, making it ideal for evaluating direct image sharpness, it fails to characterize how information from extended object features is distributed or structured for computational recovery\cite{Bertero1998Introduction}. As a result, an optical system optimized solely for geometric sharpness may inadvertently act as an information bottleneck, underperforming in a full imaging pipeline.

Meanwhile, significant advancements have been made to improve image restoration performance at the computational stage. Advanced physics-based and learning-based reconstruction methods have shown a remarkable ability to mitigate optical blur and noise. This has paved the way for end-to-end design strategies\cite{Wang2022dO,Yang2024Curriculum,Xu2025Design,Bao2025Component,Chen2023Computational}, where optical parameters and reconstruction networks are jointly optimized. By leveraging training data to maximize task-specific performance, these approaches treat the optical front-end and digital back-end as a truly unified, coupled system.

However, end-to-end design strategies are often inherently tied to specific training datasets and reconstruction architectures, which can bias the resulting optical configuration. This data-driven coupling often obscures physical interpretation of the optical system and raises concerns regarding generalization and robustness, particularly when operational conditions deviate from the training domain. These challenges underscore the necessity for optical design objectives that explicitly incorporate computational imaging requirements while remaining independent of specific reconstruction algorithms or object distributions. This leads to a fundamental question: how should an optical system be designed when image formation is no longer the final stage, but rather the first step in a computational imaging pipeline?

Estimation theory provides a natural framework for addressing this question\cite{Bertero1998Introduction,O1998Information,Chao2016Fisher,Kabuli2026Designing,Kay2013Fundamentals}. In particular, Fisher information quantifies how effectively an object's features are encoded during the imaging process, establishing a fundamental limit on reconstruction accuracy via the Cramér–Rao Bound. This perspective suggests that optical design objectives can be formulated based on information encoding efficiency rather than purely on geometric image sharpness. Motivated by this, we propose a recovery-oriented optical design approach for reconstruction-aware imaging. Rather than minimizing traditional aberrations, the proposed method optimizes the optical system to improve inversion stability, ensuring that the captured information remains reliably recoverable. This is achieved through a design criterion derived from the Fisher information matrix. Crucially, the proposed approach remains agnostic to object statistics and training data, offering a physically interpretable extension of classical design that is compatible with a diverse array of reconstruction techniques.

\section{Imaging Model and Estimation-Theoretic Background}
  \label{sc:background}

\subsection{Imaging Model}
In this work, we investigate optical imaging systems where sensor measurements are reconstructed via restoration algorithms. To characterize the relationship between optical design and reconstruction performance, we employ a matrix-based linear operator model. This framework facilitates a Fisher-information-based analysis of incoherent imaging systems under small-signal regimes.

To this end, we discretize the object plane into finite spatial elements and represent the sensor data as a vector. This formulation treats the image formation process as a linear mapping between the object and the sensor measurements, which is a commonly adopted mathematical framework in computational imaging and estimation theory. While inherently linked to the traditional point spread function (PSF) convolution model, this matrix representation provides a versatile framework for modeling general optical systems, including those with space-variant PSFs, and for evaluating reconstruction performance in the presence of noise.

Let $\bf{x}$ and  $\bf{y}$ denote the discretized object intensity distribution and the corresponding sensor measurements, respectively. By representing these distributions as column vectors, the image formation process can be formulated as

\begin{equation}
{\bf{y}} = {\bf{Hx}} + {\bf{n}}
\end{equation}

\noindent where $\bf{H}$ represents the imaging operator determined by the optical system's response, and $\bf{n}$ denotes the measurement noise. In this framework, the operator $\bf{H}$ encapsulates all optical effects, ranging from defocus to aberrations that induce blur and distortion; meanwhile, the noise term accounts for sensor noise as well as other stochastic perturbations. This linear representation serves as a versatile abstraction for investigating image restoration. Whether employing physics-based inversion or learning-based methods, the attainable reconstruction fidelity is fundamentally dictated by the spectral properties of the imaging operator $\bf{H}$.

\subsection{Fisher Information and Estimation-Theoretic Bounds}
We employ estimation theory to characterize the relationship between optical design and reconstruction performance. Assuming additive white Gaussian noise with covariance $\sigma^2\bf{I}$, the Fisher information matrix associated with the estimation of $\bf{x}$ is defined as

\begin{equation}
{\cal I} = \frac{1}{{{\sigma ^2}}}{{\bf{H}}^{\rm{T}}}{\bf{H}}
\end{equation}

\noindent The Fisher information matrix quantifies the mapping sensitivity between object perturbations and sensor data, providing a rigorous metric for the reliability of feature estimation.

A fundamental result of estimation theory is the Cramér–Rao bound (CRB), which prescribes a lower bound on the covariance of any unbiased estimator of $\bf{x}$. In practice, however, imaging operators in physical optical systems are often ill-conditioned, making the direct inversion of $\bf{H}^{\rm{T}}\bf{H}$ numerically unstable. To mitigate this, a regularized formulation is typically adopted, which admits a statistical interpretation as a Bayesian Cramér–Rao bound (BCRB) under a diffuse isotropic prior. The corresponding bound is proportional to

\begin{equation}
{({{\bf{H}}^{\rm{T}}}{\bf{H}} + \varepsilon {\bf{I}})^{ - 1}}
\end{equation}

\noindent where $\varepsilon$ is a regularization parameter that ensures numerical stability.

\subsection{Trace-Inverse Design Criterion}

Building upon these considerations, we adopt a trace-inverse objective derived from the Fisher information matrix as a reconstruction-aware optical design criterion. Specifically, we aim to minimize

\begin{equation}
trace\left( {{{({{\bf{H}}^{\rm{T}}}{\bf{H}} + \varepsilon {\bf{I}})}^{ - 1}}} \right)
\end{equation}

\noindent which corresponds to the total estimation variance across all object degrees of freedom within the regularized framework. By minimizing this sum, we effectively mitigate error amplification during the inversion process, ensuring robust and high-fidelity recovery even in noise-limited regimes. For convenience, we denote this estimation-theoretic criterion as the Recovery-Oriented Design (ROD) objective throughout the remainder of this work.

Unlike geometric metrics that focus on localized image sharpness, the trace-inverse criterion explicitly penalizes the weakly observable modes of the imaging operator, thereby promoting balanced information transfer across the spatial frequency spectrum. Consequently, it favors optical configurations that are better conditioned for computational restoration, even when such designs exhibit a slight degradation in raw imaging fidelity as measured by traditional benchmarks. Importantly, this criterion is dictated solely by the physical system and noise statistics, independent of any data-driven training. It thus provides a physically interpretable and object-independent extension of conventional optical design objectives for modern computational imaging systems.

\section{Optical System Design and Optimization}
  \label{sec:Optical System Design and Optimization}
This section describes the optical system under investigation and the optimization procedures employed to generate both RMS-optimized and recovery-oriented designs. To ensure a rigorous and unbiased comparison, both designs utilize an identical optical configuration, share the same degrees of freedom, and are subject to the same design constraints, differing solely in their respective optimization objective.

\subsection{Optical Configuration}

The evaluation system employed in this study is a dual-element separated aspheric lens. The optical configuration is selected as a representative yet non-trivial baseline, providing sufficient design degrees of freedom while maintaining computational tractability. All optical surfaces are rotationally symmetric even-order aspheres; notably, no diffractive or coded apertures are incorporated.

To isolate the effect of the optimization objective, both designs are subject to identical system specifications and constraints, including focal length, f-number, field of view, and sensor specifications. The primary design parameters are summarized in Table \ref{tab:design-parameters}.

\begin{table}[htbp]
\caption{Specifications of the dual separated aspheric imaging lens}
  \label{tab:design-parameters}
  \centering
\begin{tabularx}{0.9\linewidth}{>{\centering\arraybackslash}X >{\centering\arraybackslash}X}
\hline
Parameters & Specification \\
\hline
Focal Length (mm) & 50 \\
Field of View ($^\circ$) & 38.8 \\
F-Number & 2.0 \\
Wavelength (nm) & 486, 588, 656 \\
Sensor Size (mm) & 24.93 $\times$ 24.93 \\
\hline
\end{tabularx}
\end{table}

\subsection{Optimization Objectives}
  \label{sec:Optimization Objectives}
This subsection describes the optimization objectives and numerical procedures employed to realize the RMS-optimized and the ROD optical systems. While both designs are optimized under identical system constraints and design degrees of freedom, they differ fundamentally in their respective objective function formulations. To ensure a rigorous and unbiased comparison, consistent regularization terms and optimization settings are maintained across both design processes.

\subsubsection{RMS-based Optical Design Objective}

For the RMS-optimized design, the merit function is the RMS spot radius evaluated across multiple field points. At each field, rays spanning specified spectral range are traced, and the RMS spot radius is computed relative to the paraxial image coordinates (or the chief ray intersection) as follows:

\begin{equation}
RMS = \sqrt {\frac{1}{N}\sum\limits_{i = 1}^N {\left[ {{{\left( {{x_i} - {x_p}} \right)}^2} + {{\left( {{y_i} - {y_p}} \right)}^2}} \right]} }
\end{equation}

\noindent where $(x_i, y_i)$ denote the intersection coordinates of the $i$-th ray on the image plane, $(x_p, y_p)$ represent the ideal image coordinates predicted by paraxial optics for the corresponding field, and $N$ is the total number of rays traced across all wavelengths traced for the field. 

In addition to the RMS spot metric, several regularization terms are incorporated to ensure manufacturability and system feasibility. These terms impose penalties on center and edge thickness, as well as other geometric constraints, to preclude physically non-realizable solutions. Consequently, The total RMS-based merit function is formulated as a weighted sum of the mean RMS spot radius over all sampled field points and the regularization terms:

\begin{equation}
  \label{eq:RMS_loss}
{{\cal L}_{{\rm{RMS}}}}({\bm{\theta }}) = {w_{{\rm{RMS}}}}\;\frac{1}{{|F|}}\sum\limits_{f \in F} {{\rm{RM}}{{\rm{S}}_f}}  + \sum\limits_{k = 1}^K {{w_k}} {\mkern 1mu} {P_k}({\bm{\theta }})
\end{equation}

\noindent In Eq. \eqref{eq:RMS_loss}, $\bm{\theta}$ denotes the vector of optical design parameters, including surface profile coefficients and inter-element spacings. The set $F$ represents the sampled field points, while the second term incorporates a set of regularization penalties $\{P_k\}$, each scaled by a corresponding weight $w_k$. 

\subsubsection{Construction of the Imaging Operator and ROD Objective}

In the recovery-oriented design framework, the optical system is optimized using an estimation-theoretic objective derived from the imaging operator $\bf{H}$, as introduced in Section \ref{sc:background}. The operator $\bf{H}$ characterizes the linear mapping between the discretized object and sensor measurements and is constructed numerically via ray tracing.

To compute $\bf{H}$, the object plane is first discretized into a finite set of spatial basis elements. For each element, a bundle of rays is launched with directions sampled uniformly over the entrance pupil, ensuring that each ray represents an equivalent fraction of the optical throughput. Upon tracing these rays through the optical system, their intersections with the sensor plane are recorded. The contribution of each ray to the discrete sensor pixels is then accumulated using an interpolation-based weighting kernel, thereby forming a single column of the imaging operator $\bf{H}$. Iterating this procedure over all object elements yields the complete system matrix.

For polychromatic imaging, this characterization is performed independently for each wavelength specified in Table \ref{tab:design-parameters}. Specifically, for the three primary spectral lines $\lambda_F$, $\lambda_d$, $\lambda_C$, we construct three wavelength-dependent imaging operators ${\bf{H}}_{\rm{F}}$, ${\bf{H}}_{\rm{d}}$, ${\bf{H}}_{\rm{C}}$, are constructed. On the sensor side, the responses corresponding to each wavelength are treated as distinct spectral channels. The trace-inverse metric is then evaluated for each wavelength channel individually, and the composite objective is defined as the arithmetic mean across the three wavelength channels.

Utilizing these constructed imaging operator, the ROD objective is formulated based on the trace-inverse criterion derived from the regularized Fisher information matrix. Specifically, the optimization aims to minimize

\begin{equation}
  \label{eq:ROD_loss}
	{{\cal L}_{{\rm{ROD}}}}({\bm{\theta }}) = {w_{{\rm{ROD}}}}\sum\limits_{\lambda \in \{\mathrm{F,d,C}\}} {{\mathop{\rm trace}\nolimits} \left[ {{{\left( {{{\bf{H}}_\lambda }{{({\bm{\theta }})}^{\rm{T}}}{{\bf{H}}_\lambda }({\bm{\theta }}) + \varepsilon I} \right)}^{ - 1}}} \right]}  + \sum\limits_{k = 1}^K {{w_k}} {\mkern 1mu} {P_k}({\bm{\theta }})
\end{equation}

\noindent where $w_{{\rm{ROD}}}$ is a scaling factor calibrated to ensure that the primary objective term is commensurate in magnitude with that in Eq. \eqref{eq:RMS_loss}, while $\varepsilon$ is a regularization parameter introduced to ensure numerical stability. Throughout our experiments, we set $\varepsilon = 10^{-3}$ relative to the normalized intensity scale in $[0, 1]$. The regularization penalties are identical to those in Eq. \eqref{eq:RMS_loss}, ensuring that both RMS-based and ROD-based frameworks are subject to the same geometric and manufacturability constraints. As discussed in Section 2, this criterion seeks to minimize the total estimation variance across all object degrees of freedom, thereby promoting a stable and well-conditioned inversion.

Since direct matrix inverse in Eq. \eqref{eq:ROD_loss} is computationally prohibitive for large-scale systems, the trace-inverse term is efficiently approximated via a stochastic Lanczos method, which enables scalable estimation of the trace without explicit matrix inversion.

\subsubsection{Differentiable Ray Tracing and Optimization Strategy}

Both optimization problems are addressed using a differentiable ray tracing framework\cite{Wang2022dO,Yang2024Curriculum} that enables automatic differentiation of the objective function with respect to optical surface parameters. Ray tracing, imaging operator construction, and objective evaluation are implemented within a unified computational graph, allowing gradients to be computed efficiently via backpropagation.

To handle the large number of rays required for both RMS evaluation and imaging operator construction, the entire pipeline is implemented in a GPU-accelerated environment. Ray tracing, sensor response accumulation, and optimization objective evaluations are parallelized at the ray level, significantly reducing computational cost and facilitating large-scale optimization. Given the tractable dimensionality of the design space and the smoothness of the objective functions, the limited-memory Broyden–Fletcher–Goldfarb–Shanno (L-BFGS) algorithm is employed, ensuring rapid convergence and a consistent optimization strategy across both frameworks.

\subsection{Optimization Results}

To evaluate the impact of different optimization objectives under strictly controlled conditions, a common initial optical configuration is optimized using two distinct merit functions: the conventional RMS-based objective and the proposed ROD objective. Both optimization routines originate from an identical quasi-planar configuration, where surface radii of curvature are initially prescribed with large values. This ensures a neutral baseline that avoids pre-shaping the wavefront with strong prior aberrations.

\begin{figure}[htbp]
\centering\includegraphics[width=13.28cm]{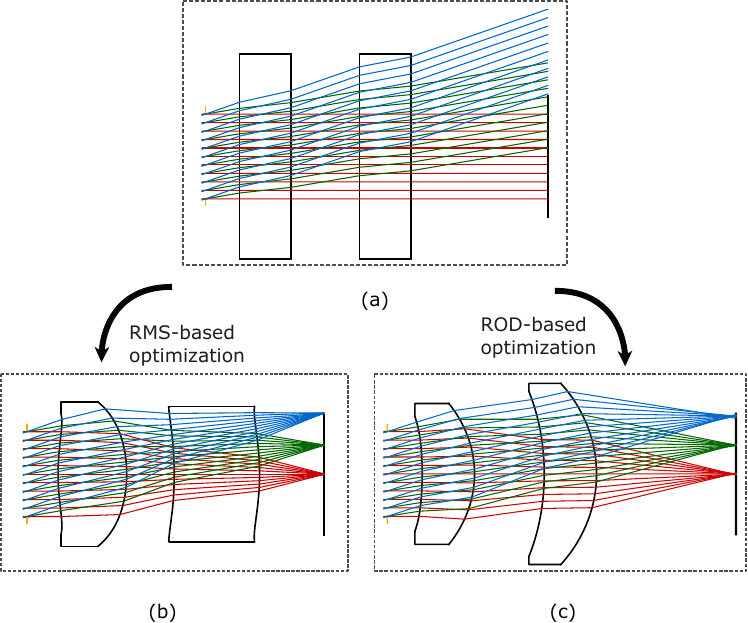}
\caption{Optical layouts: (a) Initial configuration; (b) RMS-based design; (c) ROD-based design.}
\label{fig:layout}
\end{figure}

The design variables encompass the radii of curvature for all four refractive surfaces, the even-order aspheric coefficients (from 2nd to 14th order) for each surface profile, and all axial dimensions, including lens center thicknesses, air spacings between elements, and the back focal distance. Thus, both optimization routines explore an identical design space with the same degrees of freedom.

Throughout the process, stringent geometric and manufacturability constraints are enforced uniformly using the shared penalty terms detailed in Section \ref{sec:Optimization Objectives}. Upon convergence, the penalty values for both the RMS-based and ROD-based systems are zero, confirming that both solutions strictly adhere to the prescribed physical boundaries.

Figure~\ref{fig:layout} illustrates the optical layouts of the initial configuration and the final designs obtained using the two optimization objectives. Despite sharing the same system specifications and constraints, the resulting geometries diverge, a structural disparity arising exclusively from the choice of the primary merit function. To evaluate raw imaging performance, spot diagrams are generated for four representative normalized field points of 0, 0.4, 0.7, and 1. For each field, rays across the full spectral range are traced, with their image-plane intersections plotted relative to the field centroids.

\begin{figure}[htbp]
\centering\includegraphics[width=13.28cm]{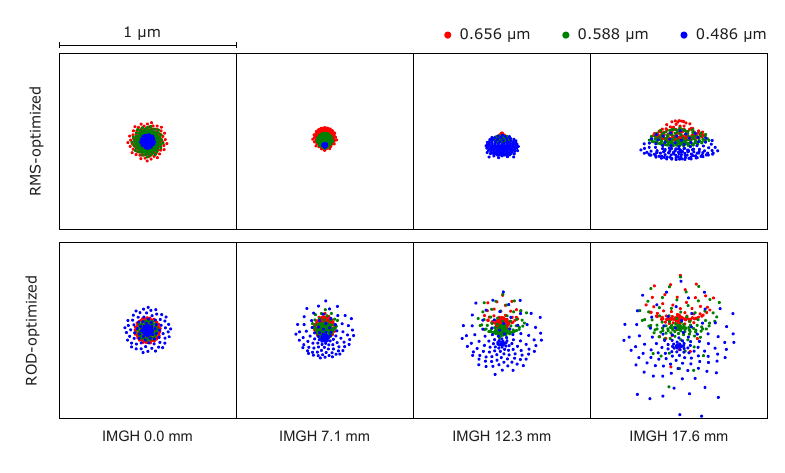}
\caption{Spot diagrams of the RMS-optimized and ROD-optimized systems at normalized field positions 0, 0.4, 0.7, and 1. Each row corresponds to one design.}
\label{fig:spot}
\end{figure}

As expected, the RMS-optimized configuration achieves significantly smaller spot radii across most field points, consistent with its objective of minimizing geometric spot size. In contrast, the ROD-optimized design shows the fundamental shift in the optimization paradigm: rather than pursuing localized sharpness, the ROD objective prioritizes the numerical conditioning of the imaging operator to facilitate robust information transfer. The impact of these structural trade-offs on computational reconstruction fidelity is evaluated in the subsequent section.

\section{Results and Reconstruction-Aware Comparison}
To compare the system-level behavior of the RMS-optimized and recovery-oriented optical designs, we evaluate both systems across the complete imaging pipeline, including direct optical rendering, physics-based inversion, and model-based neural network restoration. Two standard resolution targets are used in this study: the NBS 1963 resolution chart, which emphasizes high-frequency line patterns, and the ISO 12233 chart, which contains slanted edges and mixed spatial-frequency structures. These two targets provide complementary structural characteristics, enabling a comprehensive evaluation of both fine-line resolution and edge fidelity under different reconstruction strategies.

For each optical system, the object images are first rendered through the differentiable ray-tracing framework described in Section \ref{sec:Optical System Design and Optimization} to generate direct sensor images. These rendered images are then processed using two reconstruction approaches: a linear Tikhonov inversion method and a model-based neural network restoration framework. The optical designs remain fixed during reconstruction. To simulate realistic measurement conditions, additive white Gaussian noise is introduced to the rendered sensor images prior to reconstruction. The noise standard deviation is set to $\sigma = 10^{-4}$, relative to normalized image intensity in the range $[0, 1]$. This noise level is kept identical for both optical systems and for all reconstruction experiments to ensure fair comparison.

Reconstruction quality is evaluated using peak signal-to-noise ratio (PSNR) and structural similarity index (SSIM), both computed with respect to the original ground-truth images. These metrics are reported for direct rendering results as well as for the reconstructed images obtained from the two restoration methods.

\subsection{Direct Imaging Performance}

\begin{figure}[htbp]
\centering\includegraphics[width=13.28cm]{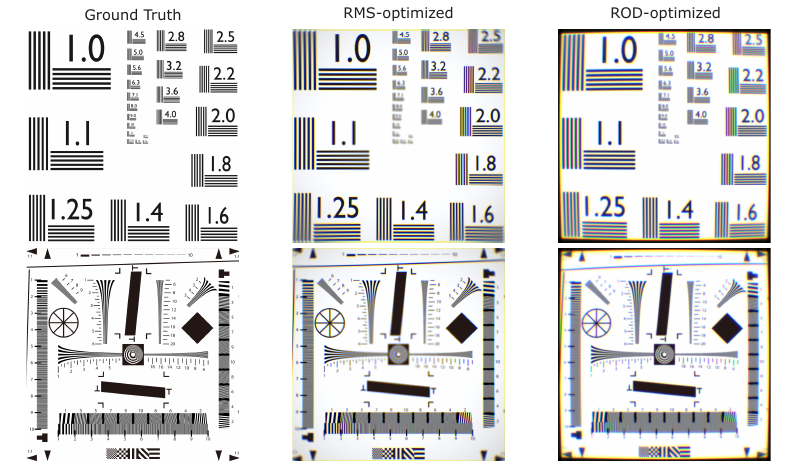}
\caption{Direct imaging comparison between the RMS-optimized and recovery-oriented optical systems. The left column shows the ground-truth test charts (NBS 1963 and ISO 12233). The middle and right columns present the corresponding rendered sensor images produced by the RMS-optimized lens and ROD-optimized lens, respectively.}
\label{fig:direct imaging}
\end{figure}

We first compare the direct imaging results produced by the two optical systems. Figure~\ref{fig:direct imaging} presents the original test charts alongside the rendered sensor images produced by the RMS-optimized and recovery-oriented designs.

Although the RMS-optimized design yields higher apparent sharpness in low-frequency and high-contrast structures, it exhibits significant degradation in high-spatial-frequency details. High-frequency line pairs undergo more severe aliasing, and certain small-scale structures exhibit contrast exaggeration without preserving correct geometric separation. This behavior suggests that certain spatial-frequency components are transferred with reduced stability, potentially limiting subsequent inversion performance.

In contrast, the ROD-optimized system produces images with reduced apparent sharpness; however, high-frequency line pairs remain structurally distinguishable, and directional features exhibit greater consistency. On both test charts, small characters, fine grids, and slanted edges retain coherent geometric structure without premature frequency collapse. The edge transitions are more monotonic and stable, suggesting improved suitability for subsequent inversion or restoration. Rather than maximizing peak sharpness, the ROD design preserves a more balanced distribution of Fisher information across spatial frequencies, preventing certain modes from collapsing toward near-zero sensitivity.

An additional difference is observed in geometric distortion. The ROD-optimized system exhibits noticeably stronger distortion, particularly near the image periphery, where straight features appear slightly warped relative to the ground truth. This reflects an inherent trade-off introduced by the recovery-oriented optimization criterion, where spatial uniformity and distortion are not explicitly constrained in the trace-inverse objective. Consequently, the design prioritizes invertibility and information preservation over geometric fidelity. Despite the increased distortion, the structural integrity of fine details remains intact, indicating that the recovered information content is better preserved even if geometric mapping deviates more strongly from ideal projection.

\subsection{Reconstruction via Tikhonov Regularization}

To analyze inversion behavior under a linear reconstruction framework, we apply Tikhonov regularization to the noisy rendered images. Given the linear imaging model ${\bf{y}} = {\bf{Hx}} + {\bf{n}}$, the reconstructed image is obtained by solving the following regularized least-squares problem:

\begin{equation}
  {{\bf{\hat x}}_\lambda } = \arg \mathop {\min }\limits_{\bf{x}} \left\| {{\bf{Hx}} - {\bf{y}}} \right\|_2^2 + \gamma \left\| {\bf{x}} \right\|_2^2
\end{equation}

\noindent where $\gamma$ is the regularization parameter controlling the trade-off between data fidelity and noise suppression. The corresponding closed-form solution is

\begin{equation}
  {{\bf{\hat x}}_\lambda } = {\left( {{{\bf{H}}^{\bf{T}}}{\bf{H}} + \gamma {\bf{I}}} \right)^{ - 1}}{{\bf{H}}^{\bf{T}}}{\bf{y}}.
\end{equation}

In our implementation, the linear system is solved iteratively using a conjugate gradient method without explicitly forming dense matrices. For each optical system, $\gamma$ is varied over multiple orders of magnitude, and the reconstruction quality is quantified using PSNR and SSIM. The optimal reconstruction for each system is determined by the value of $\gamma$ that maximizes PSNR.

\begin{figure}
\centering\includegraphics[width=13.28cm]{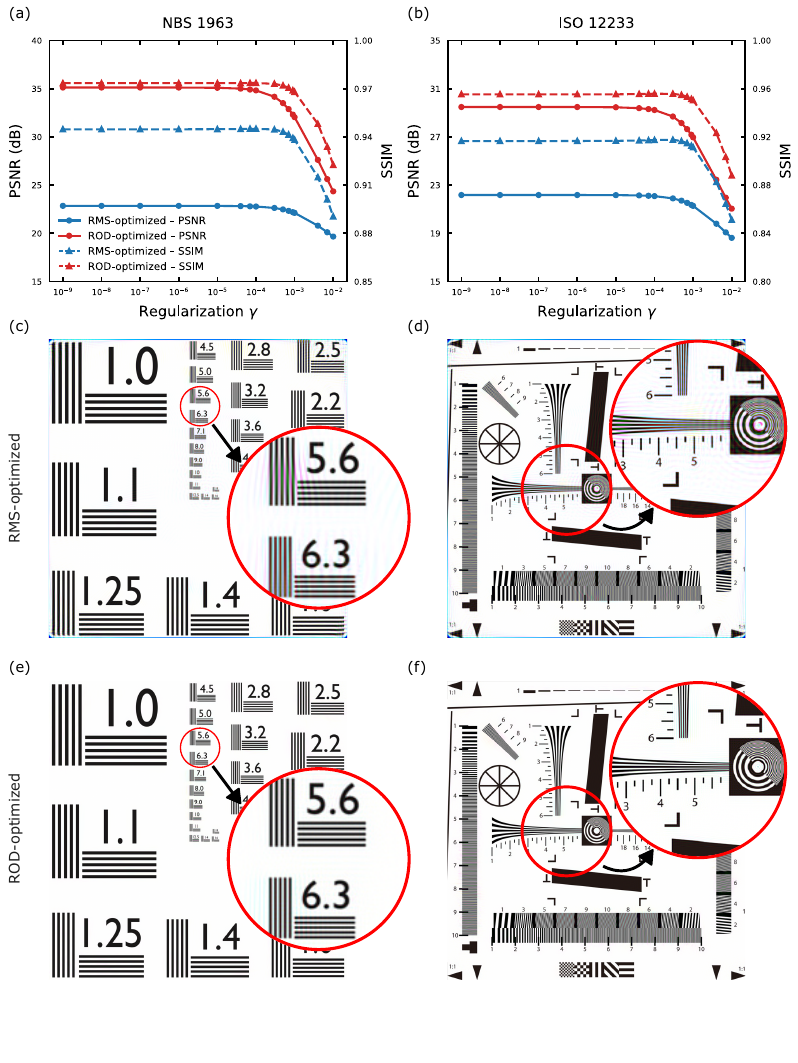}
\caption{Regularized linear inversion results and parameter sensitivity analysis. (a–b) PSNR (solid lines, left vertical axis) and SSIM (dashed lines, right vertical axis) as functions of $log_{10}(\gamma)$ for the NBS 1963 chart (left) and ISO 12233 chart (right). (c–f) The corresponding Tikhonov reconstruction results of the NBS 1963 and ISO 12233 charts using individually optimized regularization parameters for the RMS-optimized and recovery-oriented optical systems.  }
\label{fig:tikhonov}
\end{figure}

Figure~\ref{fig:tikhonov} presents both the reconstructed images and the corresponding parameter sensitivity curves. The upper portion of the figure illustrates PSNR (left vertical axis) and SSIM (right vertical axis) as functions of $log_{10}(\gamma)$. The lower portion of the figure shows the reconstruction results for the NBS 1963 and ISO 12233 charts using the individually optimized $\gamma$ for each optical system.

The quantitative results in panels (a) and (b) of Figure~\ref{fig:tikhonov} reveal a consistent performance gap between the two optimization strategies across a wide range of the regularization parameter $\gamma$. For both the NBS 1963 and ISO 12233 datasets, the ROD-optimized system (red curves) achieves higher PSNR and SSIM values than the RMS-optimized system (blue curves) over most of the tested parameter range. While both systems exhibit a characteristic plateau for small values of $\gamma$, followed by a performance decline as regularization becomes excessive (above approximately $10^{-4}$), the ROD-optimized curves remain systematically above those of the RMS-based design. This robust performance profile suggests that the ROD objective effectively minimizes the condition number of the imaging operators ${\bf{H}}$, leading to enhanced structural preservation and increased robustness under regularized inversion. Thus, the system is more amenable to deconvolution across a broad range of regularization strengths.

The reconstructed images further corroborate these quantitative observations. For the NBS 1963 chart shown in panels (c) and (e), the RMS-optimized reconstruction (c) exhibits visible ringing and color artifacts in the magnified regions around the “5.6” and “6.3” line groups. High-contrast edges display oscillatory behavior and partial contrast instability, indicating increased sensitivity to inversion-induced error amplification. In contrast, the ROD-optimized reconstruction (e) exhibits superior suppression of artifact and more stable separation between adjacent line pairs. The boundaries appear more geometrically coherent, and spurious oscillations are noticeably mitigated.

A similar trend is observed in the ISO 12233 results shown in panels (d) and (f). The RMS-optimized reconstruction (d) exhibits noticeable aliasing and Moiré-like artifacts in the Siemens star and high-frequency slanted-line regions, where fine structures become irregular and unstable under inversion. By comparison, the ROD-optimized reconstruction (f) maintains clearer structural organization in these dense patterns, with lines remaining distinguishable closer to the center of the target and reduced artifact amplification near high-contrast transitions.

These results demonstrate that the recovery-oriented optical design leads to a forward operator that is inherently more stable under regularized inversion. The improvement is not attributable to selective parameter tuning but reflects intrinsic differences in reconstruction conditioning induced by the two optimization strategies.

\subsection{Learning-Based Image Restoration}

To further evaluate whether the reconstruction advantage of ROD persists under nonlinear inverse modeling, we adopt a model-based deep learning (MoDL) framework. Unlike purely feedforward image-to-image networks, MoDL explicitly incorporates the forward imaging operator into the network structure through unrolled optimization, thereby enforcing data consistency with respect to the physical model.

Given the linear imaging model ${\bf{y}} = {\bf{Hx}} + {\bf{n}}$, the MoDL framework aims to solve a regularized problem of the form

\begin{equation}
  {\bf{\hat x}} = \arg {\min _{\bf{x}}}\left\| {{\bf{Hx}} - {\bf{y}}} \right\|_2^2\; + \beta \left\| {{\bf{x}} - {{\cal D}_\theta }({\bf{x}})} \right\|_2^2
\end{equation}

\noindent where ${\cal D}_{\theta}(\cdot)$ denotes a learnable denoiser and $\beta$ controls the strength of the learned prior. Rather than solving this problem directly, the network unrolls a fixed number of alternating updates. At iteration k, the denoiser first produces

\begin{equation}
  {{\bf{z}}_k} = {{\cal D}_\theta }({{\bf{x}}_k})
\end{equation}

\noindent followed by a data-consistency update obtained by solving the linear system

\begin{equation}
  ({{\bf{H}}^{\bf{T}}}{\bf{H}} + \beta {\bf{I}}){{\bf{x}}_{k + 1}} = {{\bf{H}}^{\bf{T}}}{\bf{y}} + \beta {{\bf{z}}_k}
\end{equation}

\begin{figure}
\centering\includegraphics[width=13cm]{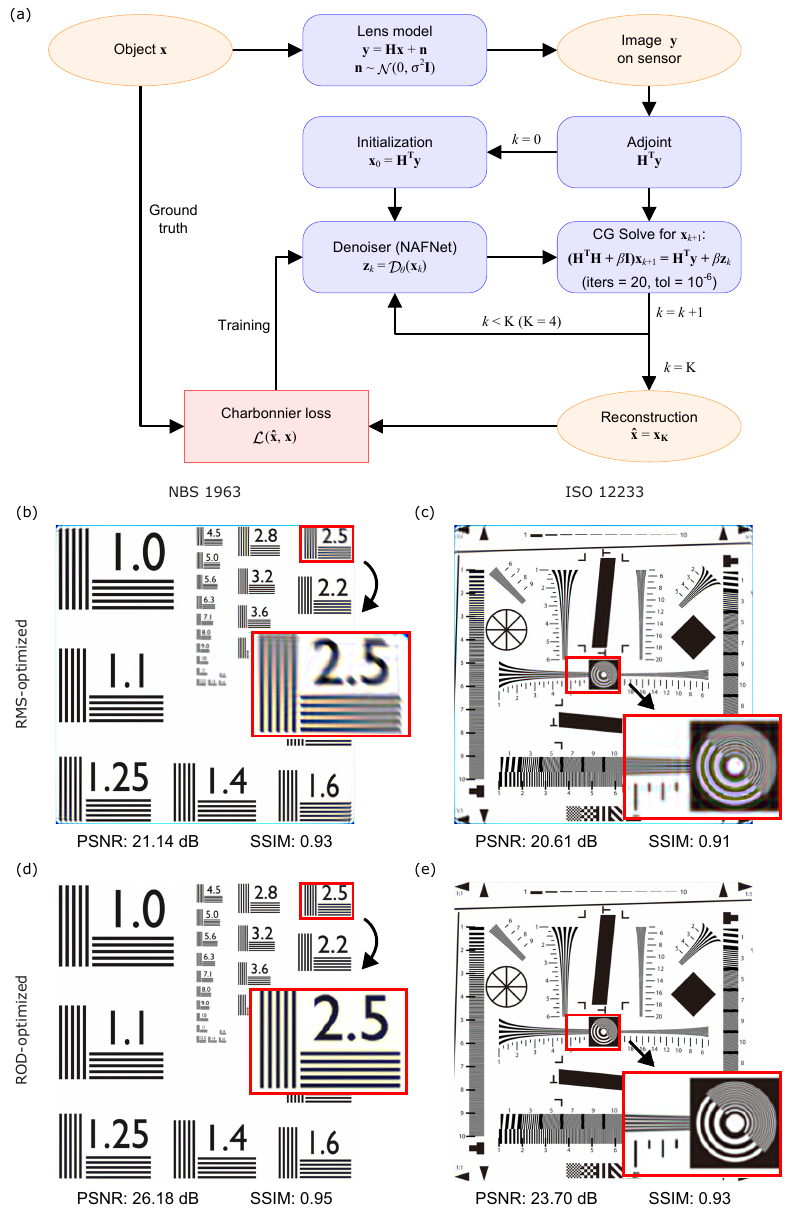}
\caption{Results of model-based deep learning (MoDL) reconstruction. (a) Unrolled MoDL architecture integrating the fixed forward operator H, learned denoiser ${\cal D}_{\theta}$ (NAFNet), and conjugate-gradient-based data-consistency updates. (b-e) Reconstruction results for the NBS 1963 and ISO 12233 charts using identical network architecture and training protocol for both optical systems. Quantitative metrics (PSNR/SSIM) are reported for each reconstruction. The recovery-oriented design consistently achieves higher PSNR and SSIM, while also preserving fine structures and reducing artifacts in high-frequency regions.}
\label{fig:MoDL}
\end{figure}

In our implementation, this system is solved using conjugate gradients with 20 inner iterations and tolerance $10^{-6}$. The unrolling depth is set to $K = 4$, and the data-consistency weight is fixed at $\beta = 0.05$ for both systems.

The denoiser ${\cal D}_{\theta}$ is implemented using a lightweight NAFNet architecture with width 24, a single middle block, encoder block configuration $[1, 1, 1, 8]$, and decoder configuration $[1, 1, 1, 1]$. The same network architecture and hyperparameters are used for both the RMS-optimized and recovery-oriented optical systems to ensure a fair comparison. During training, the optical system remains fixed; gradients are propagated only through the denoiser parameters, while the forward and adjoint operators are treated as deterministic physics modules.

To reduce memory consumption, we employ truncated backpropagation through the unrolled iterations: gradients are propagated only through the final outer iteration, while the preceding iterations are executed without storing intermediate computational graphs. This strategy substantially reduces memory cost while preserving stable training behavior.

Figure~\ref{fig:MoDL} illustrates the MoDL reconstruction pipeline together with the restoration results. Quantitative metrics (PSNR/SSIM) are also reported in each subfigure, providing a direct comparison between the two optical designs. Overall, the recovery-oriented design consistently achieves higher PSNR and SSIM values than the RMS-optimized system, confirming the visual improvements.

For the NBS 1963 resolution chart, the RMS-optimized system exhibits residual ringing and color artifacts in the magnified “2.5” region. In contrast, the recovery-oriented design produces improved edge definition and reduced artifact amplification. Line pairs remain more distinctly separated, and transitions between dark and bright regions appear more geometrically consistent. These qualitative improvements are consistent with the observed gains in PSNR and SSIM, indicating better preservation of high-frequency information.

For the ISO 12233 chart, the qualitative behavior under MoDL is largely consistent with the trends observed in the Tikhonov inversion. Although nonlinear reconstruction suppresses certain artifacts in both systems, the recovery-oriented design continues to exhibit improved structural coherence near the resolution limit. Fine gratings, circular features, and edge transitions are better preserved, whereas the RMS-optimized system exhibits partial frequency collapse and local contrast distortions. These visual improvements are further supported by consistently higher PSNR and SSIM values for the recovery-oriented design, indicating that the enhanced structural fidelity translates into measurable reconstruction gains.

The consistency across both linear and deep-learning-based regimes underscores the algorithm-agnostic nature of the proposed optimization, indicating that the gain originates from the enhanced information conditioning of the imaging operator $\bf{H}$. In summary, while RMS optimization prioritizes perceptual sharpness in the raw image, ROD optimization produces an image that is better conditioned for computational reconstruction, thereby maximizing the fidelity of the final output regardless of the restoration backend.

\section{Conclusion and Outlook}
In this work, we introduced a recovery-oriented optical design framework based on an estimation-theoretic criterion derived from the Fisher information matrix. Instead of optimizing conventional image-domain sharpness metrics, the proposed method directly improves the conditioning of the forward imaging operator by minimizing the trace-inverse of a regularized information matrix. This shifts the design objective from visual appearance at the sensor plane toward computational recoverability.

By embedding a differentiable physical model into the optimization loop, the resulting optical system is explicitly designed with downstream reconstruction in mind. The Fisher-information-based objective enhances the sensitivity and stability of the imaging operator itself, leading to a more balanced spatial-frequency response and improved invertibility. As demonstrated through both linear regularized inversion and model-based nonlinear reconstruction, the performance gain is consistent across distinct reconstruction paradigms, indicating that the improvement originates from intrinsic operator properties rather than from algorithm-specific effects.

It is useful to contrast this strategy with end-to-end optical–neural co-design approaches, in which optical parameters and reconstruction networks are jointly optimized over a particular training dataset. While such methods can achieve strong empirical performance, the learned optical configuration may become implicitly coupled to the statistics of the training scenes and the chosen network architecture. In contrast, the recovery-oriented framework proposed here optimizes the physical forward operator itself, producing an imaging system whose reconstruction advantage persists across different inversion algorithms without relying on dataset-specific adaptation.

Overall, this work highlights a complementary pathway for integrating optics and computation. Rather than fully coupling optical parameters to a specific neural network, the proposed approach emphasizes principled operator-level optimization grounded in estimation theory. As computational imaging continues to evolve, combining such physics-driven conditioning criteria with modern reconstruction algorithms may provide a robust and generalizable foundation for future imaging system design.

\begin{backmatter}
\bmsection{Funding}
This work is supported by National Natural Science Foundation of China (12274156); National Key R\&D Program of China (2023YFC2414700); Science, Technology and Innovation Commission of Shenzhen Municipality, China (JCYJ20250604191014019).

\bmsection{Disclosures}
The authors declare no conflicts of interest.

\bmsection{Data Availability Statement}
Data underlying the results presented in this paper are not publicly available at this time but may be obtained from the authors upon reasonable request.

\end{backmatter}


\bibliography{sample}

\end{document}